\documentclass[aps,
twocolumn,
reprint,
superscriptaddress,
amsmath,
amssymb,
]{revtex4-2}
\usepackage[left]{lineno}
\usepackage[english]{babel}
\usepackage{graphicx}
\usepackage[colorlinks=true, allcolors=blue]{hyperref}
\usepackage{times}
\usepackage{multirow}
\usepackage{bm}
\usepackage{braket}
\usepackage{booktabs}
\usepackage{listings}
\usepackage[normalem]{ulem}

\begin{document}

\title{Learning unified control of intrinsic nonlinear spin dynamics in atomic qudits for magnetometry}

\author{C.-Z. Cao}
\thanks{These authors contributed equally to this work.}
\affiliation{College of Physics \& Key Laboratory of Aerospace Information Perception and Physics, MIIT, Nanjing University of Aeronautics and Astronautics, Nanjing 211106, China}

\author{J.-Z. Han}
\thanks{These authors contributed equally to this work.}
\affiliation{China Mobile (Suzhou) Software Technology Co., Ltd., Suzhou 215163, China}

\author{M. Xiong}
\affiliation{College of Physics \& Key Laboratory of Aerospace Information Perception and Physics, MIIT, Nanjing University of Aeronautics and Astronautics, Nanjing 211106, China}

\author{M. Deng}
\affiliation{Yichang Testing Technique R\&D Institute, Specialized Weak Magnetic Metering Station of NDM, Yichang 443003, China}

\author{L. Wang}
\affiliation{Shanghai Institute of Optics and Fine Mechanics, Chinese Academy of Sciences, Shanghai 201800, China}

\author{Q. Liu}
\email{qiliu26@nju.edu.cn}
\affiliation{National Laboratory of Solid State Microstructures and School of Physics, Collaborative Innovation Center of Advanced Microstructures, Nanjing University, Nanjing 210093, People’s Republic of China}

\author{X. Lv}
\email{xudonglv@siom.ac.cn}
\affiliation{Shanghai Institute of Optics and Fine Mechanics, Chinese Academy of Sciences, Shanghai 201800, China}

\author{M. Xue}
\email{mxue@nuaa.edu.cn}
\affiliation{College of Physics \& Key Laboratory of Aerospace Information Perception and Physics, MIIT, Nanjing University of Aeronautics and Astronautics, Nanjing 211106, China}
\affiliation{Center for the Cross-disciplinary Research of Space Science and Quantum-technologies (CROSS-Q), Nanjing University of Aeronautics and Astronautics, Nanjing 211106, China}

\begin{abstract}
Generating and preserving metrologically useful quantum states is a central challenge in quantum-enhanced metrology.
In low-field atomic magnetometry with multilevel atoms, the nonlinear Zeeman (NLZ) effect is both a resource and a limitation.
It can generate internal spin squeezing within a single atomic qudit, but under fixed readout it also rotates and distorts the measurement-relevant quadrature, limiting the usable metrological gain.
The problem is further complicated by the time dependence of both the squeezing axis and the nonlinear evolution itself.
Here we show that reinforcement learning can transform NLZ dynamics from a source of readout degradation into a sustained metrological resource. 
Using only experimentally accessible low-order spin moments, a trained agent identifies a unified control policy for this class of intrinsically nonlinear sensing dynamics.
We illustrate the approach in the \(f=21/2\) manifold of \(^{161}\mathrm{Dy}\), where the learned policy rapidly prepares strongly squeezed internal states and stabilizes more than \(4\,\mathrm{dB}\) of fixed-axis spin squeezing under continuous NLZ evolution.
Including state-preparation overhead, the learned protocol yields a single-atom magnetic-field sensitivity of \(13.9\,\mathrm{pT}/\sqrt{\mathrm{Hz}}\), approximately \(3\,\mathrm{dB}\) beyond the standard quantum limit.
Our results establish learning-based control as an experimentally feasible route for converting unavoidable intrinsic nonlinear dynamics in multilevel atomic sensors into operational metrological advantage.
\end{abstract}

\maketitle

\section{Introduction}

\begin{figure*}
\centering
\includegraphics[width=1.99\columnwidth]{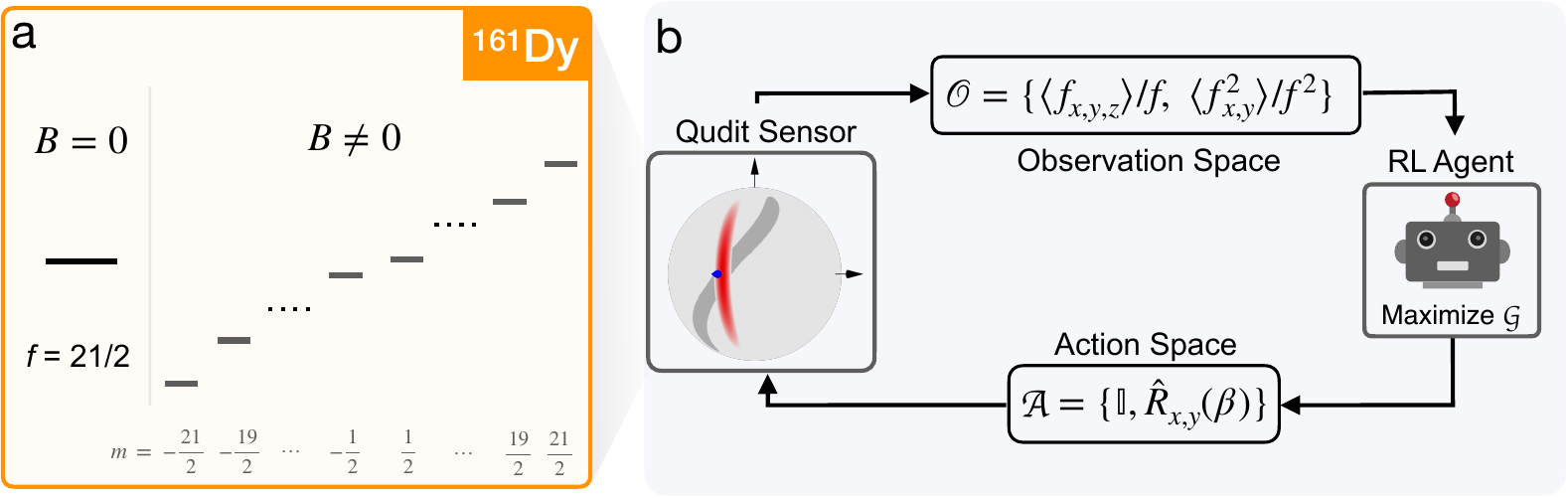}
\caption{
\textbf{Reinforcement-learning framework for controlling intrinsic spin dynamics in an atomic qudit.}
(a) A single \(^{161}\mathrm{Dy}\) atom in the \(f=21/2\) hyperfine manifold forms a \((2f+1)\)-dimensional qudit. A magnetic field induces an effective quadratic Zeeman contribution, producing intrinsic nonlinear spin dynamics.
(b) The qudit evolves under interleaved nonlinear evolution and transverse control rotations. The red shaded region on the Bloch sphere denotes a metrologically useful spin-squeezed state, while the gray deformed region denotes its distortion under continued nonlinear evolution. At each step, the agent receives a reduced observation \(\mathbf{o}_k\in\mathcal O\), selects an action \(a_k\in\mathcal A\), and maximizes the cumulative reward \(\mathcal G\) to learn a unified policy for squeezing generation and stabilization.
}
\label{fig:rlscheme}
\end{figure*}

Quantum sensing is a central paradigm for precision measurement.
Among its leading platforms, atomic magnetometry has broad applications in fundamental physics~\cite{jackson2017constraints,wang2020single,fedderke2021earth,fedderke2021search,arza2022earth,wei23ultasensitive,su24new,felix25levitated,cong25rmp}, biomagnetic detection~\cite{sander2012magnetoencephalography,kamada2015human,he2019high,zhang2020recording}, and magnetic navigation~\cite{canciani2016absolute,canciani2017airborne,gnadt2022machine,zhang2023heading,xiao2023femtotesla,lei2025sensitivity}.
Quantum-enhanced metrology can further improve sensitivity beyond the standard quantum limit (SQL) by exploiting nonclassical resources such as squeezing and entanglement~\cite{pezze2018quantum,huang24review,MONTENEGRO20251}. 
In multilevel atoms, quantum enhancement need not rely on interparticle entanglement, but can instead be encoded within the internal degrees of freedom of a single spin-$f$ system~\cite{Fernholz08spin}, equivalently viewed as a symmetric ensemble of $2f$ spin-$1/2$ constituents~\cite{Kurucz10multilevel,satoor21partitioning}.

Recent experiments have established the internal Hilbert space of multilevel atoms as a metrological resource~\cite{chalopin2018quantum,evrard2019enhanced,satoor21partitioning,Hemmer21squeezing}. Atomic qudits can host nonclassical internal states, including non-Gaussian and cat-like states generated by controllable nonlinear dynamics such as light-induced tensor shifts, yielding enhanced magnetic sensitivity and even near-Heisenberg-limited performance under suitable control~\cite{yang2025minute,zhang25cooperative}. In these settings, however, the nonlinearity is externally engineered and can be switched or shaped, so that state preparation can be cleanly separated from metrological interrogation.

In realistic low-field sensing regimes, by contrast, the relevant nonlinearity is often intrinsic. The quadratic Hamiltonian $\propto \hat{f}_z^2$, arising from the nonlinear Zeeman (NLZ) effect, is the single-atom counterpart of the familiar one-axis-twisting (OAT) interaction in many-body spin systems~\cite{kitagawa1993squeezed}. It naturally generates internal spin squeezing, but simultaneously rotates the squeezing axis and drives overtwisting on experimentally relevant timescales~\cite{chalopin2018quantum}. As a result, the same dynamics that create nonclassical correlations can also degrade the metrological gain accessible under fixed readout~\cite{acosta2006nonlinear,budker2007optical,jensen2009cancellation}. This difficulty is most acute when the nonlinearity is generated by the same external field that carries the signal, so that nonlinear distortion accompanies signal accumulation throughout the sensing process. A prototypical example is Earth-field-range magnetometry, where the NLZ effect arises directly from the ambient magnetic field being measured and has motivated a range of compensation, spin-locking, and staged-control approaches that either suppress its influence or recover part of the metrological gain through fixed dynamical-decoupling pulse structures and separate squeezing-generation and stabilization stages~\cite{seltzer2007synchronous,wasilewski2010quantum,bao2018suppression,lee2021heading,shaniv2019quadrupole,yang2021coherence,bao2022all,yang2025quantum,djlj-pnlk}. 
Because the same field simultaneously drives both useful signal encoding and unwanted nonlinear evolution, the optimal control is intrinsically nonseparable: squeezing generation and stabilization must be interleaved in time to preserve the measurement-relevant quadrature.

To address this intrinsically nonseparable control problem, we adopt a  reinforcement-learning framework~\cite{sutton1998reinforcement}. Reinforcement learning is well suited to this setting because the control protocol is naturally sequential, while the impact of individual actions on fixed-readout metrological performance is not directly transparent in the high-dimensional Hilbert space~\cite{Metz23self,meng2023machine,duan2025concurrent,Duncan25taming,bukov2026reinforcement}. 
RL has been applied to a broad range of quantum control and sensing problems, including state preparation, parameter estimation, and multiqubit disentanglement~\cite{bukov2018reinforcement,xiao2022parameter,tashev2026reinforcement}.
Unlike structure-constrained dynamical-decoupling strategies based on predefined pulse structures and staged protocols~\cite{degen2017quantum}, it can interact directly with the dynamics during training and discover pulse sequences tailored to the underlying system.
Specifically, we consider resonant transverse radio-frequency rotations interleaved with intrinsic nonlinear dynamics, and train an agent to select control actions using only experimentally accessible low-order spin moments. 
Guided by a reward that captures both overall squeezing and measurement-relevant fixed-axis performance, the learned unified policy rapidly generates strong internal spin squeezing while stabilizing the measurement-relevant quadrature under continuous evolution. The learned policy also reveals a simple and physically interpretable pulse structure that suppresses overtwisting while preserving useful quantum correlations, offering an experimentally feasible route to operational metrological advantage in intrinsically nonlinear quantum sensors.

\section{Model and Unified Reinforcement-Learning Framework}

\subsection{Intrinsic nonlinear dynamics and squeezing metrics}
We consider an atomic qudit encoded in the \(d=2f+1\) Zeeman states \(|f,m\rangle\) (\(m=-f,\ldots,f\)) within a single hyperfine manifold in the presence of a magnetic field $B$, which defines the quantization \(z\) axis. As illustrated in Fig.~\ref{fig:rlscheme}(a), we take the \(^{161}\mathrm{Dy}\) manifold with \(f=21/2\) as a representative example~\cite{lu2012quantum}, where the applied field lifts the Zeeman degeneracy and induces an effective quadratic nonlinear Zeeman contribution. In the weak-field regime, the resulting hyperfine--Zeeman structure is described by the effective single-atom Hamiltonian (\(\hbar=1\))
\begin{equation}
\hat H_{B} \simeq \Omega_L \hat f_z + \chi \hat f_z^2,
\end{equation}
where \(\{\hat f_x,\hat f_y,\hat f_z\}\) are the internal spin operators, \(\Omega_L\) is the Larmor frequency, and \(\chi\) denotes the quadratic Zeeman effect (QZE) coefficient.
In the rotating frame that removes the Larmor precession, the dynamics reduce to
\begin{equation}
\hat H_{\rm QZE} = \chi \hat f_z^2,
\label{eq:HQZE}
\end{equation}
which is formally equivalent to the OAT Hamiltonian~\cite{kitagawa1993squeezed}, 
here acting on the internal $(2f{+}1)$-dimensional manifold of a single spin.

Starting from a coherent spin state (CSS) polarized along $x$,
$|\psi_0\rangle = |f, m_x{=}f\rangle$, the always-on quadratic term shears the
spin distribution in the plane transverse to the mean spin, thereby generating
a spin-squeezed state, shown schematically as the red shaded region on the Bloch sphere in Fig.~\ref{fig:rlscheme}(b). Continued nonlinear evolution, however, rotates the squeezing axis and induces overtwisting, illustrated by the gray deformed region, rapidly degrading the metrological gain under fixed readout. This competition between squeezing generation and distortion motivates active control.

To quantify metrologically useful squeezing during the evolution of a single spin-$f$ relative to a CSS,
we use the Wineland parameter~\cite{wineland1992spin}
\begin{equation}
\xi^2=\frac{2f\,\min_{\perp}[(\Delta f_\perp)^2]}
{|\langle\mathbf{\hat f}\rangle|^2},
\label{eq:xiR}
\end{equation}
which directly quantifies the phase uncertainty relative to the SQL under the optimal linear readout.
It is widely used to assess metrological enhancement in spin-squeezing experiments~\cite{greve2022entanglement,mao2023quantum,robinson2024direct}.
For a CSS, $\xi^2=1$, while $\xi^2<1$ indicates sub-SQL phase uncertainty.
Since our sensing protocol uses a fixed readout axis, we further introduce
\begin{equation}
\xi_y^2=\frac{2f\,(\Delta f_y)^2}{|\langle \hat f_x\rangle|^2},
\label{eq:xiW}
\end{equation}
which can be regarded as the fixed-axis counterpart of the Wineland parameter.
It follows from the same error-propagation relation, but with the measurement direction fixed to the $y$ axis rather than optimized over transverse directions to minimize the spin fluctuations.
Consequently, $\xi_y^2$ quantifies the phase uncertainty relative to the SQL for the measurement of spin component $\hat f_y$ considered here.

\begin{figure*}[!hbt]
\centering
 \includegraphics[width=2.0\columnwidth]{Fig2.pdf}
\caption{\textbf{Learned pulse protocol and squeezing dynamics.}
(a) Pulse sequence selected by the RL agent. Colored bars denote discrete transverse rotations applied at each control step.
(b) Time evolution of the Wineland squeezing parameter \(\xi^2(t)\) (red solid) and the fixed-axis squeezing parameter \(\xi_y^2(t)\) (green solid) versus the dimensionless time \(\chi t\). Gray solid and dotted curves denote $\xi^2(t)$ of the QZE (OAT) evolution and the effective TACT benchmark, respectively. The vertical gray line near \(\chi t\simeq 0.15\) marks the reward-defined switching time \(t_r\). The inset Wigner distributions show representative snapshots at selected times \(t_i\); the colored shaded regions on each Bloch sphere represent the spin Wigner distributions of the atomic state at the labeled times.}
 \label{fig:pulseseq}
\end{figure*}

\subsection{Reinforcement-learning control framework}\label{subsec2_2}
The overall control loop is summarized in Fig.~\ref{fig:rlscheme}(b). 
The single-atom spin-$f$ manifold is controlled by a transverse radio-frequency 
drive. In the same rotating frame used for Eq.~(\ref{eq:HQZE}), and under the
rotating-wave approximation, the dynamics are described by $\hat{H}_{\rm QZE} +\hat{H}_{\rm ctrl}$ with $\hat{H}_{\mathrm{ctrl}}
=  \lambda(\hat{f}_x\cos\phi+\hat{f}_y\sin\phi)$,
where $\lambda$ is the Rabi frequency and $\phi$ sets the rotation axis
in the transverse $xy$ plane. 
For numerical optimization, we discretize the control into a sequence of
piecewise-constant pulses. The total duration $T$ is divided into $N_t$
intervals of length $\delta t=T/N_t$. In the $k$th interval, the control is
fully specified by the rotation angle $\beta_k=\int_{k\delta t}^{(k+1)\delta t}\lambda\,dt$.
We denote the corresponding instantaneous transverse rotation by
$\hat{R}_{\mu}(\beta_k)=e^{-i\beta_k \hat f_{\mu}}$,
where $\mu$ labels the rotation axis determined by the phase $\phi$. 
A Rabi frequency $\lambda/2\pi$ on the order of tens of kilohertz is sufficient to justify the instantaneous rotation approximation. 

Each control step then consists of a spin rotation followed by evolution
under the QZE Hamiltonian,
\begin{equation}
|\psi_{k+1}\rangle
= e^{-i\chi \delta t \hat f_z^2}\hat{R}_{\mu}(\beta_k)|\psi_k\rangle.
\label{eq:update}
\end{equation}
This stroboscopic update is the dynamical backbone of the learning loop shown in
Fig.~\ref{fig:rlscheme}(b), where transverse control rotations are interleaved
with the intrinsic nonlinear evolution.

The agent selects actions from a discrete set of experimentally natural
rotations,
\begin{equation}
\mathcal{A}=\left\{\mathbb{I},\,\hat R_{\mu}(\beta)\; \big|\; \mu\in\{x,y\},\ \beta\in\{\pm\tfrac{\pi}{2}, \pm\tfrac{\pi}{3}, \pm\tfrac{\pi}{4}\}\right\},
\label{eq:action}
\end{equation}
where $\mathbb{I}$ denotes no control pulse.
While this symmetric parametrization is convenient, the two rotation channels
play distinct physical roles.
Pairs of opposite-angle $\hat R_y$ rotations enable symmetric toggling-frame control that compensates the nonlinear rotation induced by the QZE~\cite{liu2011spin,chen2019extreme}. By contrast, $\hat R_x$ pulses primarily act as corrective rotations that reorient the squeezed state toward the measurement-relevant axis.
For this purpose, positive rotation angles are already sufficient in practice, as confirmed by the optimized pulse sequences discussed below.

As indicated by the observation-space block in Fig.~\ref{fig:rlscheme}(b), 
the agent does not access the full quantum state. 
Instead, at each step it receives a low-dimensional observation vector
$\mathbf{o}_k\in\mathcal O$ ($\mathcal O \subset \mathbb R^5 $) constructed from experimentally accessible spin moments,
\begin{equation}
\mathbf{o}_k =
(
{\langle\hat f_x\rangle_k}/{f},
{\langle\hat f_y\rangle_k}/{f},
{\langle\hat f_z\rangle_k}/{f},
{\langle\hat f_x^2\rangle_k}/{f^2},
{\langle\hat f_y^2\rangle_k}/{f^2}
).\nonumber
\label{eq:obs}
\end{equation}
These moments can be obtained from measurements of spin projections and their
variances~\cite{chalopin2018quantum,evrard2019enhanced,satoor21partitioning,yu2025schrodinger}, 
and suffice to track the mean-spin direction and transverse fluctuations 
relevant to squeezing and readout~\cite{cox2016deterministic,hosten2016measurement,pedrozo2020entanglement,robinson2024direct}.
Importantly, the components of $\mathbf{o}_k$ are normalized by $f$, 
rendering the input representation independent of the Hilbert-space dimension 
and facilitating adaptation to different spin manifolds.

The control objective is unified throughout training and is encoded in a scalar reward $r_k$ constructed from the physically relevant squeezing metrics $\xi^2$ and $\xi_y^2$. The agent is trained to maximize the cumulative return
\begin{equation}
\mathcal{G}=\sum_{k=0}^{N_t-1} \gamma^k_{\mathrm{RL}}r_k,
\end{equation}
where $\gamma_{\mathrm{RL}} \in (0,1)$ is the discount factor. This unified objective is designed to favor both rapid squeezing generation and the preservation of measurement-relevant squeezing under fixed readout.

To implement this objective, we adopt a curriculum-style reward with a state-dependent switching time defined by the first-hitting time
\begin{equation}
k_{\star} \equiv \inf\{k \ge 0:\, \xi^2(k)\le \xi_{\rm ref}^2\},
\end{equation}
namely, the first step at which the Wineland squeezing parameter reaches a prescribed threshold.
The reference value $\xi_{\rm ref}^2$ is fixed throughout training and chosen as the optimal squeezing benchmark attainable under ideal two-axis counter-twisting (TACT) dynamics~\cite{kitagawa1993squeezed,liu2011spin}.
Here, the TACT squeezing value serves as a reference threshold for the RL optimization. 
Its role is to guide the RL strategy to reach the TACT-level squeezing, after which the optimization objective switches to the stabilization of the fixed-axis squeezing.
The reward function is then defined as
\begin{equation}
r_k =
\begin{cases}
\displaystyle
\log \xi^2(k-1) - \log \xi^2(k) - c(a_k),
& k < k_\star, \\[6pt]
\displaystyle
\zeta e^{-\kappa k_\star} - c(a_k),
& k = k_\star, \\[6pt]
\displaystyle
-\alpha\log \xi_y^2(k) - c(a_k),
& k > k_\star,
\end{cases}
\label{eq:reward}
\end{equation}
where the first term rewards progress toward stronger squeezing, the second provides a one-time bonus favoring earlier threshold crossing, and the third promotes post-threshold preservation of the fixed-axis squeezing $\xi_y^2$, which is most relevant for metrological performance under our readout scheme. The cost term $c(a_k)$ penalizes pulse usage, with $a_k\in\mathcal A \backslash \{ \mathbb{I} \}$. Here $\zeta$ sets the bonus scale, $\kappa$ its temporal decay with the hitting time $k_\star$, and $\alpha$ the weight assigned to post-threshold stabilization (see Appendix~\ref{app:MethodA}).

We optimize the cumulative return using Proximal Policy Optimization (PPO)~\cite{schulman2017proximal,wang2020truly,gu2021proximal}, 
an Actor--Critic policy-gradient method~\cite{konda1999actor,grondman2012survey}. 
The policy network maps the low-order moment observation \(\mathbf{o}_k\in\mathcal O\) to a control action, 
while the value network estimates the expected return. 
This model-free architecture enables stable learning without an explicit dynamical model~\cite{porotti2019coherent,guo2021faster,yao2021reinforcement,porotti2022deep,li2025robust,bukov2026reinforcement}, 
allowing the agent to discover a unified policy for squeezing generation and long-time stabilization.

\section{Learned control strategy, stabilization mechanism, and metrological gain}
\label{sec2_3}

\subsection{Learned protocol structure}
\label{subsec2_3}

We now analyze the control protocol learned by the agent. Although the reward is unified, the learned policy develops a clear two-stage structure, as seen in the pulse sequence of Fig.~\ref{fig:pulseseq}(a): an initial stage that rapidly generates squeezing, followed by a stabilization stage that preserves the measurement-relevant quadrature. The corresponding evolution of $\xi^2$ and $\xi_y^2$, together with spin-Wigner snapshots at selected times, is shown in Fig.~\ref{fig:pulseseq}(b).

During the initial stage, the learned sequence consists of three paired 
$\hat R_y(\pm\pi/2)$ operations. 
These paired pulses suppress QZE-induced overtwisting and accelerate squeezing generation, 
allowing the system to reach a minimum value of 
$\xi^2=-8.19\,\mathrm{dB}$ near \(t_3\). 
This is comparable to the optimal TACT benchmark of 
$-8.07\,\mathrm{dB}$ and surpasses the QZE (OAT) benchmark of 
$-7.17\,\mathrm{dB}$. 
The first two inset Bloch spheres in Fig.~\ref{fig:pulseseq}(b) visualize this process: 
the initial CSS and the state at \(t_1\) are shown by the gray and red shadings on the first sphere, 
while the state at \(t_2\) and an intermediate state are shown by the red and light-red shadings on the second sphere.

After the squeezing reaches the reference threshold $\xi_{\rm ref}^2$, 
the agent applies an isolated $\hat R_x(\pi/3)$ pulse near \(t_3\), 
rotating the state toward an orientation where the fixed-axis squeezing $\xi_y^2$ is close to its minimum. 
Rather than entering the alternating $\hat R_y(\pm\pi/2)$ cycle directly, 
the protocol inserts an additional $\hat R_y(-\pi/4)$ pulse. 
This transition is shown on the third inset Bloch sphere in Fig.~\ref{fig:pulseseq}(b): 
the state near \(t_3\) is indicated in gray, 
the post-$\hat R_x(\pi/3)$ state in yellow, 
and the post-$\hat R_y(-\pi/4)$ state near \(t_4\) in cyan.
The intermediate $\hat R_y(-\pi/4)$ rotation is essential for stabilization. 
Without it, direct entry into the alternating $\hat R_y(\pm\pi/2)$ cycle would drive one of the two visited states to a much larger $\xi_y^2$, 
producing large oscillations in the measurement-relevant squeezing. 
The learned protocol instead balances $\xi_y^2$ between the two alternating states 
and pre-aligns the state for phase accumulation under the same pulse pattern 
(see Appendix~\ref{app:phasesensry}). 
This stabilized cycle is shown on the fourth inset Bloch sphere in Fig.~\ref{fig:pulseseq}(b), 
with $\xi_y^2(t_5)=-5.11\,\mathrm{dB}$ and $\xi_y^2(t_6)=-4.03\,\mathrm{dB}$. 
Thus, $\xi_y^2$ remains metrologically useful and exhibits only weak bounded oscillations 
despite the continued action of the QZE term.

\begin{figure}[!b]
\centering
\includegraphics[width=0.92\columnwidth]{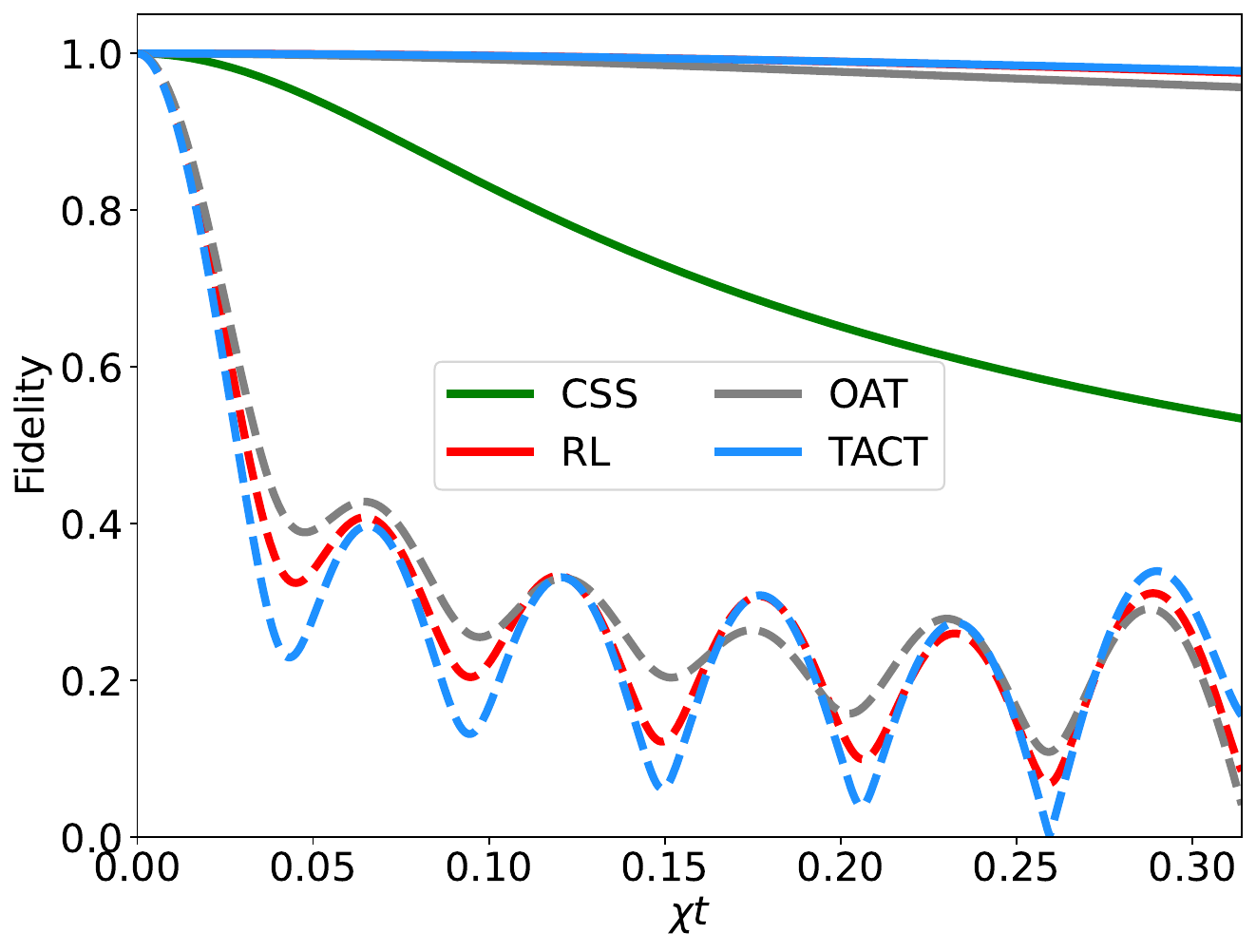}
\caption{\textbf{Fidelity evidence for the stabilization mechanism.}
Fidelity evolution for representative reference states. Solid and dashed curves denote evolution under \(\hat f_y^2\) and \(\hat f_z^2\), respectively. The reference states are the coherent spin state \(\lvert f,m_x{=}f\rangle\), the RL-generated state after the $\hat R_x(\pi/3)$ pulse in Fig.~\ref{fig:pulseseq}, and the OAT and TACT states evaluated at their respective minima of \(\xi^2\). For the OAT and TACT states, an additional \(\hat R_x\) rotation is applied so that \(\xi_y^2=\xi^2\).}
\label{fig:fy_fz}
\end{figure}

\begin{figure*}[!hbt]
\centering
\includegraphics[width=1.98\columnwidth]{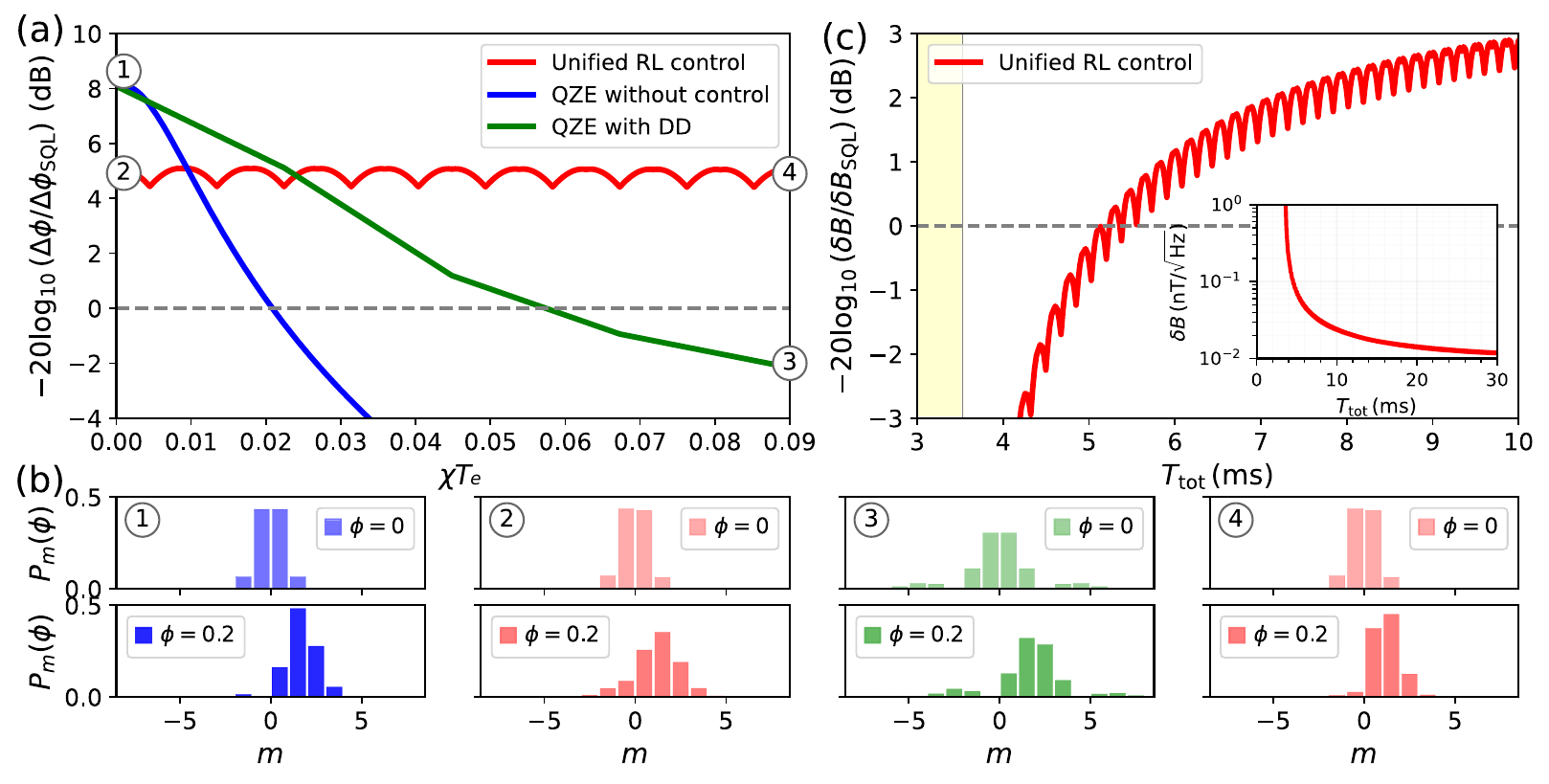}
\caption{\textbf{Metrological analysis of the unified RL control strategy.}
(a) Phase sensitivity relative to the SQL, expressed in dB, as a function of the dimensionless interrogation time $\chi T_e$.
The red curve corresponds to the unified RL control starting from the stabilized state at time $t_4$ in Fig.~\ref{fig:pulseseq}, the blue curve corresponds to bare QZE evolution of the TACT-optimal squeezed
state without control, and the green curve corresponds to the DD protocol.
The gray dashed line denotes the SQL. The SQL is defined by the CSS phase sensitivity $\Delta \phi_{\mathrm{SQL}} = 1/\sqrt{2f}$.
(b) Readout distributions $P_m(\phi)$ for representative probe states and control protocols. 
Panels~\textcircled{\scriptsize 1} and~\textcircled{\scriptsize 2} show the distributions before ($\phi=0$, top) and after ($\phi=0.2$, bottom) phase encoding for the initial states used in the blue and red curves of panel (a), respectively. 
Panels~\textcircled{\scriptsize 3} and~\textcircled{\scriptsize 4} show the corresponding distributions at $\chi t\approx 0.09$ for the green and red curves in panel (a), respectively.
(c) Single-atom magnetic-field sensitivity relative to the SQL, expressed in dB, as a function of the total protocol time $T_{\mathrm{tot}}$. 
The shaded region indicates the preparation time $T_{\mathrm p}$, and the inset shows the absolute sensitivity $\delta B(T_{\mathrm{tot}})$ in $\mathrm{nT}/\sqrt{\mathrm{Hz}}$.
}
\label{fig:metrology}
\end{figure*}

The learned stabilization strategy differs qualitatively from the 
\(\hat R_x\)-pulse dynamic-decoupling (DD) protocol of 
Ref.~\cite{yang2025quantum}. 
Instead, the RL agent identifies a stabilization mechanism based on alternating 
transverse \(\hat R_y\) rotations. 
For one elementary cycle of duration \(dt=2\delta t\),
\begin{equation}
\hat U_{\rm cyc}
=
e^{-i\chi\delta t\hat{f}_z^2}\,
\hat R_y(-\pi/2)\,
e^{-i\chi\delta t\hat{f}_z^2}\,
\hat R_y(\pi/2).
\label{eq:ucyc}
\end{equation}
The second free-evolution interval is transformed by the intermediate 
\(\hat R_y\) rotations as
$\hat R_y(-\pi/2)\,
e^{-i\chi\delta t\hat{f}_z^2}\,
\hat R_y(\pi/2)
=
e^{-i\chi\delta t\hat{f}_x^2}$,
since 
$\hat R_y(-\pi/2)\hat{f}_z^2 \hat R_y(\pi/2)=\hat{f}_x^2$. 
Thus, to leading order,
$\hat U_{\rm cyc}
=
e^{-i\chi\delta t\hat{f}_z^2}
e^{-i\chi\delta t\hat{f}_x^2}
=
e^{ \{-i\chi \delta t(\hat{f}_z^2+\hat{f}_x^2)
+\frac{1}{2}(\chi\delta t)^2[\hat{f}_x^2,\hat{f}_z^2]
+O((\chi\delta t)^3)\} }$.
For the parameters used here, \(\chi\delta t\sim10^{-3}\), 
so the higher-order terms are negligible. 
Using 
\(\hat f_x^2+\hat f_z^2=f(f+1)-\hat f_y^2\), 
the effective generator is therefore proportional to 
\(-\hat f_y^2\), up to an irrelevant constant. 
The alternating \(\hat R_y(\pm\pi/2)\) pulses thus redirect the nonlinear evolution 
from direct \(\hat f_z^2\) twisting to an effective \(\hat f_y^2\)-type dynamics, 
which is more compatible with preserving the fixed-axis squeezing \(\xi_y^2\). 
Since this effective evolution acts weakly on states with small 
\(\Delta  f_y\), the measurement-relevant quadrature remains comparatively stable.

This interpretation is supported by the fidelity analysis in Fig.~\ref{fig:fy_fz}. Under \(\hat f_z^2\) evolution, states with smaller \(\xi_y^2\) show faster fidelity decay, indicating greater susceptibility to QZE-induced distortion. By contrast, evolution under \(\hat f_y^2\) preserves both the RL- and TACT-generated squeezed states much more effectively, consistent with the stabilization of the measurement-relevant squeezed quadrature.

\subsection{Phase sensitivity and metrological gain}

We now evaluate the metrological performance of the learned protocol under fixed projective readout of $\hat f_y$.
In the stabilization regime, the periodic control cycle maintains an approximately constant fixed-axis squeezing $\xi_y^2$, 
thereby providing a suitable operating point for sensing.

During the sensing stage, we consider the local estimation of a weak signal $B_{\mathrm{sig}}$ around a known bias field $B_0$. For $B_{\mathrm{sig}}\ll B_0$, the signal-induced variation of $\chi$ is negligible, and $\chi$ is determined solely by $B_0$. The signal is encoded over a duration $T_e$ as a phase $\phi=\gamma B_{\mathrm{sig}} T_{\mathrm e}$, where $\gamma$ is the gyromagnetic ratio.
The encoded state is
\begin{equation}
|\psi_{\mathrm{enc}}\rangle
=
\prod_{k=1}^{N_e}
\Big[
e^{-i\left(
\chi \hat{f}_z^2+\frac{\phi}{T_{\mathrm e}}\hat{f}_z
\right)\delta t}
\,\hat R_y((-1)^k\pi/2)
\Big]
|\psi ({t_4})\rangle ,
\label{eq:psiencode}
\end{equation}
where $|\psi ({t_4})\rangle$ is the stabilized state at time $t_4$ in Fig.~\ref{fig:pulseseq}, 
and $N_e=\lceil T_{\mathrm e}/\delta t \rceil$ is the number of control steps during encoding. 
After encoding, the phase is read out from the spin projection $\hat f_y$.
For small $\phi \rightarrow 0$, the phase sensitivity is quantified by
\begin{equation}
(\Delta\phi)^2
=
\frac{(\Delta f_y)^2}
{|\partial_\phi \langle \hat{f}_y \rangle|^2},
\label{eq:erroepropa}
\end{equation}
which is the standard error-propagation formula for local parameter estimation~\cite{toth2014quantum}.

Fig.~\ref{fig:metrology}(a) shows the phase sensitivity of the unified RL protocol as a function of the dimensionless encoding time \(\chi T_{\mathrm e}\). 
As references, we consider two protocols starting from the TACT-optimal squeezed state: bare QZE evolution without control and the DD protocol of Ref.~\cite{yang2025quantum} 
applied to the same initial state (see Appendix~\ref{app:xxx} for details). 
The unified RL protocol instead starts from the stabilized state \(|\psi ({t_4})\rangle\), 
whose fixed-axis squeezing \(\xi_y^2\) is weaker than that of the TACT-optimal squeezed state. 
At short encoding times, the TACT-optimal probes therefore perform slightly better.

This initial advantage is short-lived. 
As the encoding time increases and QZE-induced distortion accumulates, 
the uncontrolled TACT-optimal probe loses its advantage beyond the SQL already around 
\(\chi T_{\mathrm e}\approx 0.02\), 
while the DD protocol retains a beyond-SQL advantage only up to about 
\(\chi T_{\mathrm e}\approx 0.06\) before degrading rapidly. 
By contrast, the unified RL protocol maintains a phase-sensitivity gain of about 
\(5\,\mathrm{dB}\) over the SQL across a broad range of encoding times, with only weak residual oscillations. 
Thus, once QZE-induced distortion becomes appreciable, 
the long-time stability of the unified RL protocol outweighs the larger initial squeezing of the reference probes. 

We also compare the achieved phase sensitivity with the Heisenberg limit (HL).
A spin-$f$ qudit can be viewed as the symmetric subspace of $2f$ spin-$1/2$ constituents, for which the HL is $\Delta\phi_{\rm HL}=1/(2f)$.
The maximum phase-sensitivity gain relative to the SQL is therefore $13.2\,\mathrm{dB}$.
The approximately $5\,\mathrm{dB}$ phase-sensitivity gain maintained by the unified RL protocol remains about $8.2\,\mathrm{dB}$ away from the single-qudit HL.
The HL represents an ideal ultimate bound, whereas the present work focuses on achieving sustained beyond-SQL sensitivity under experimentally relevant constraints, including always-on nonlinear dynamics and fixed-axis readout.
It would be interesting to explore RL-compatible approaches to further narrow the gap, for instance, by leveraging extreme spin-squeezed state or time-reversal protocols~\cite{carrasco2024dicke}.

Fig.~\ref{fig:metrology}(b) shows the corresponding readout distributions
\(P_m(\phi)=|\langle f,m|\psi_{\mathrm{enc}}(\phi)\rangle|^2\).
At \(\chi T_{\mathrm e}\approx 0\), before appreciable QZE-induced distortion accumulates,
the initial probe distributions for both the TACT-optimal protocol and the unified RL protocol
remain narrow and close to Gaussian.
At \(\chi T_{\mathrm e}\approx 0.09\), however, the DD protocol shows a noticeable distortion of the readout distribution
[panel~\textcircled{\scriptsize 3} in Fig.~\ref{fig:metrology}(b)],
whereas the unified RL protocol preserves a narrow, approximately Gaussian profile
[panel~\textcircled{\scriptsize 4} in Fig.~\ref{fig:metrology}(b)].

For a realistic ${}^{161}\mathrm{Dy}$ atomic qudit in an bias Earth-field-scale magnetic field,
\(B_0 \approx50~\mu\mathrm{T}\), the relevant parameters are
\(\chi=(2\pi)\,8.112~\mathrm{Hz}\) and
\(\gamma=(2\pi)\,13.24~\mathrm{GHz/T}\) 
(see Appendix~\ref{app:hyperfine}).
To quantify performance under a finite time budget, 
we convert the phase uncertainty into a magnetic-field sensitivity. 
Neglecting the readout time, the total protocol time is
\[
T_{\mathrm{tot}}=T_{\mathrm p}+T_{\mathrm e},
\]
where \(T_{\mathrm p}\) is the fixed preparation time. 
For the unified RL protocol, \(T_{\mathrm p}\) is the time required to prepare the stabilized state 
\(|\psi(t_4)\rangle\) under QZE evolution. 
The single-atom magnetic-field sensitivity is
\begin{equation}
\delta B
=
\Delta B\,\sqrt{T_{\mathrm{tot}}}
=
\frac{\Delta\phi\,\sqrt{T_{\mathrm{tot}}}}{\gamma T_{\mathrm e}}.
\label{eq:sensitivity}
\end{equation}
As a benchmark, we use the standard quantum limit (SQL) for magnetic-field sensitivity, 
as achieved by a coherent spin state (CSS), 
evaluated in the absence of QZE and without preparation overhead, so that 
\(T_{\mathrm{tot}}=T_{\mathrm e}\):
\begin{equation}
\delta B_{\mathrm{SQL}}(T_{\mathrm{tot}})
=
\frac{1}{\gamma\sqrt{2f\,T_{\mathrm{tot}}}}.
\end{equation}

Fig.~\ref{fig:metrology}(c) shows the SQL-relative magnetic-field sensitivity
\(\delta B_{\mathrm{SQL}}/\delta B\) as a function of \(T_{\mathrm{tot}}\). 
Because of the finite preparation overhead, the unified RL protocol initially has no beyond-SQL advantage. 
After this preparation cost is amortized, it surpasses the SQL at 
\(T_{\mathrm{tot}}\approx 5.5\,\mathrm{ms}\) and continues to improve at longer times. 
At \(T_{\mathrm{tot}}=30\,\mathrm{ms}\), it reaches 
\(13.9\,\mathrm{pT}/\sqrt{\mathrm{Hz}}\), 
corresponding to an advantage of approximately \(3\,\mathrm{dB}\) beyond the SQL. 
For fixed \(\chi\), increasing \(T_{\mathrm{tot}}\) corresponds to stronger accumulated QZE effects, 
so the continued improvement demonstrates that the learned protocol preserves metrologically useful correlations over extended interrogation times.
Overall, the analysis above assumes unitary dynamics.
When experimentally relevant noise sources, such as pulse imperfections (see Appendix~\ref{app:pulsenoise}) and dephasing (see Appendix~\ref{app:dephasing}), are included, the achievable sensitivity is ultimately limited by an optimal finite total protocol time.

Although the above analysis uses \(^{161}\mathrm{Dy}\) as a representative example, 
the learned control principle is not specific to this manifold. 
Across different spin-\(f\) systems, independently trained agents recover the same qualitative organization: 
rapid squeezing generation followed by stabilization under continued QZE evolution. 
The detailed pulse sequences remain system dependent, but the underlying mechanism is robust across spin dimension. 
Detailed results for different spin-\(f\) systems are provided in Appendix~\ref{app:spinf}.

\section{Summary and outlook}

We have shown that reinforcement learning provides a unified control strategy for turning the intrinsic QZE nonlinearity of an atomic qudit into an operational metrological resource. 
Using only low-order spin moments and transverse rotations, the learned protocol integrates rapid internal spin-squeezing generation, long-time stabilization of the measurement-relevant quadrature, and signal interrogation under continuous nonlinear evolution. 
After accounting for preparation overhead, this unified control yields a net magnetic-field sensitivity beyond the SQL. 
Beyond performance improvement, the learned policy reveals an interpretable control principle: alternating $\hat R_y(\pm\pi/2)$ pulses redirect the bare $\hat f_z^2$ twisting into an effective dynamics that is less destructive to fixed-axis squeezing. 
Thus, RL supplies not merely an optimized pulse sequence, but a physically interpretable strategy for unifying state preparation, stabilization, and sensing in intrinsically nonlinear quantum systems.
Unlike previous experiments using optically engineered OAT interactions~\cite{chalopin2018quantum,evrard2019enhanced,satoor2021partitioning}, our scheme exploits the intrinsic QZE of the magnetic field.
Although the intrinsic nonlinearity is much weaker and therefore acts on a longer timescale, no additional optical field is required, thereby avoiding additional light-induced perturbations.

This viewpoint suggests several extensions. 
First, because the framework uses experimentally accessible moments and transverse rotations, it can be transferred to other high-spin atoms and collective effective-spin platforms. 
In arrays of Dy atoms, independent sensors would already provide the standard \(1/\sqrt{N}\) improvement, while dipole-dipole interactions could generate additional interatomic squeezing~\cite{bornet2023scalable,eckner2023realizing,block2024scalable}. 
Combining learned internal-qudit control with interaction-induced many-body entanglement may therefore offer a route toward stronger quantum enhancement in many-atom sensors~\cite{Kurucz10multilevel,norris2012enhanced,zhang25cooperative,hu2026enhancingcollectivespinsqueezing}. 
Platforms with spatially resolved readout could further enable parallelized and site-resolved quantum sensing~\cite{PRXQuantum.5.010311,PhysRevLett.134.193201,v5nm-xwp5}.

Second, the learning objective itself can be made more metrology-native. 
Here the reward is built from squeezing-based metrics, but future protocols could optimize the full sensing task directly. 
For signals with prior information, Bayesian posterior objectives could replace squeezing-only rewards, shifting the target from state preparation to end-to-end estimation performance~\cite{PhysRevX.11.041045,25ds-9724,ma2025adaptive}. 
Together with recent progress in coherent qudit manipulation for nonclassical state preparation and quantum algorithms~\cite{yu2025schrodinger,shi2026efficient}, these directions point to learning-based control as a general strategy for harnessing intrinsic nonlinear dynamics across multilevel quantum sensors and qudit information-processing platforms.

\section*{Acknowledgements}
M. X. was supported by the National Natural Science Foundation of China (No.~12304543), the NUAA funding (No.~1018-ILF26028), and the Quantum Science and Technology - National Science and Technology Major Project (No.~2021ZD0302100). X. L. was supported by the National Key R\&D Program of China (No.~2025YFF0515500) and the Shanghai Municipal Science and Technology Commission Strategic Frontier Special Project (No.~25DP2600100). J.-Z. H. was supported by the Natural Science Foundation of Jiangsu Province (No.~BK20250404) and the Youth Science and Technology Talent Support Program of Jiangsu Province (No.~JSTJ-2025-600) and Suzhou (No.~2025(062)). C.-Z. C. was supported by the Postgraduate Research \& Practice Innovation Program of NUAA (No.~xcxjh20252107).

\section*{DATA AVAILABILITY}
The data that support the findings of this study are publicly available in Zenodo~\cite{Cao2026Zenodo}. Additional information is available from the corresponding author upon reasonable request.

\appendix

\section*{Appendix}

\section{Ground-State Hyperfine Structure of Dy Atoms}
\label{app:hyperfine}

In the presence of an external magnetic field, the internal structure of a Dy atom is governed by the combined hyperfine interaction between the electronic and nuclear angular momenta and their Zeeman couplings to the field. Denoting the nucleus and electron angular momenta by $i$ and $j$, the full Hamiltonian is given by~\cite{Corney2003Atomic}
\begin{equation}
\begin{aligned}
\hat H
=&
\mathrm{A}\, \hat{\mathbf i}\!\cdot\!\hat{\mathbf j}+
\mathrm{B}
\frac{
\tfrac{3}{2}\hat{\mathbf i}\!\cdot\!\hat{\mathbf j}
\left(
2\hat{\mathbf i}\!\cdot\!\hat{\mathbf j}
+1-i(i+1)-j(j+1)
\right)
}
{2i(2i-1)j(2j-1)}
\\[6pt]
&+
g_j \mu_B \hat j_z B
+
g_i \mu_N \hat i_z B,
\label{eqa:hyperfine}
\end{aligned}
\end{equation}
where $h$ is the Planck constant, $\mathrm{A}$ and $\mathrm{B}$ are the magnetic-dipole and electric-quadrupole hyperfine coefficients, $g_j$ and $g_i$ are the electron and nuclear Landé g factors, $\mu_B$ and $\mu_N$ are Bohr and nuclear magnetons, $B$ is the magnetic ﬁeld
strength. In this work, the magnetic field is assumed to be aligned along the laboratory $z$ axis.

We focus on the isotope $^{161}\mathrm{Dy}$, whose ground state has $j=8$ and $i=5/2$. The hyperfine interaction splits the ground state into manifolds labeled by the total angular momentum $f$. In this work, we restrict attention to the manifold with $f=21/2$. 

In the weak-field regime, the Zeeman interaction is non-negligible but remains smaller than the hyperfine splitting. Consequently, the energy spectrum within a fixed hyperfine manifold exhibits nonlinear dependence on the magnetic quantum number $m$. After numerically diagonalizing Eq.~(\ref{eqa:hyperfine}) in the $| f,m\rangle$ basis for a given magnetic field $B$, the eigenenergies $E_m$ can be accurately fitted by a quadratic expansion,
\begin{equation}
    E_m \simeq \hbar\bigl(\Omega_L m + \chi m^2\bigr),
    \label{eqa:energy}
\end{equation}
where $\Omega_L$ is the Larmor frequency and $\chi$ is the QZE coefficient, also referred to as the NLZ coefficient or quantum-beat revival frequency.

For atoms with $j=1/2$, both $\Omega_L$ and $\chi$ can be obtained analytically from the Breit–Rabi formula~\cite{breit1931measurement}. However, for higher spin atoms such as ${}^{161}\mathrm{Dy}$ with $j>1/2$, no closed-form expression exists, and the numerical fitting is essential. For the bias magnetic field in Earth-field strengths around $B_0 = 50~\mu\mathrm T$,  we obtain $\Omega_L = (2\pi)~661.9~\mathrm{kHz}$,
$\chi = (2\pi)~8.112~\mathrm{Hz}$. The Larmor frequency satisfies $\Omega_L = \gamma B_0$, which corresponds to the gyromagnetic ratio $\gamma = (2\pi)~13.24 ~\mathrm{GHz/T}$~\cite{lu2012quantum}.

The internal dynamics are therefore well described by the Hamiltonian
\begin{equation}
\hat H
=
\hbar\bigl(
\Omega_L \hat f_z + \chi \hat f_z^2
\bigr),
\label{eqa:qzehamiltonian}
\end{equation}
where $\hat f_z$ is the $z$-component of the total angular-momentum operator for a single atom. The quadratic term $\propto \hat f_z^2$ gives rise to intrinsic OAT dynamics, which plays a central role in both the generation of spin squeezing and the QZE-induced degradation of metrological performance discussed in the main text.

\section{PPO algorithm}\label{app:MethodA}
We train the RL agent using PPO, known as an Actor-Critic algorithm. The agent is parameterized by a policy network $\pi_{\bm{\theta}}$ as an actor parameterized by $\bm{\theta}$, and a value network $V_{\bm{w}}$ as a critic parameterized by $\bm{w}$.

The system state at the \(k\)-th time step is the full quantum state \(|\psi_k\rangle\) of the atom, which evolves according to the controlled Schrödinger dynamics in Eq.~(\ref{eq:update}). Instead of accessing the full state, the agent receives an observation \(o_k\) at each discrete control step \(k\), consisting of first and second moments of the spin operators. Based on \(o_k\), the policy samples an action \(a_k\) from the discrete action set in Eq.~(\ref{eq:action}), corresponding to experimentally accessible transverse rotations. The environment then evolves the underlying quantum state from \(|\psi_k\rangle\) to \(|\psi_{k+1}\rangle\), yielding the next observation \(\mathbf{o}_{k+1}\) and a scalar reward \(r_k\) in Eq.~(\ref{eq:reward}). Transitions \((\mathbf{o}_k, a_k, r_k, \mathbf{o}_{k+1})\) are collected into a trajectory buffer and used for subsequent updates.

The value network outputs the state value \(V_{\bm{w}}(\mathbf{o}_k)\), which estimates the expected return from observation \(\mathbf{o}_k\); these value estimates are then used to compute the generalized advantage estimation (GAE),
\begin{equation}
\hat A_k
=
\sum_{l=0}^{N_t-k-1}(\gamma_{\mathrm{RL}}\hat \lambda)^l
\big[
r_{k+l}+\gamma_{\mathrm{RL}} V_{\bm{w}}(\mathbf{o}_{k+l+1})-V_{\bm{w}}(\mathbf{o}_{k+l})
\big],
\label{eq:gae}
\end{equation}
where \(\gamma_{\mathrm{RL}} \in (0,1) \) is the discount factor and \(\hat \lambda\) is a hyperparameter that controls the bias--variance trade-off.

The policy network is updated by maximizing the clipped objective function
\begin{equation}
L_{\mathrm{actor}}(\bm{\theta})
=
\mathbb{E}_k\!\left[
\min\!\left(
\eta_k(\bm{\theta})\hat A_k,\,
\mathrm{clip}(\eta_k(\bm{\theta}),1-\epsilon,1+\epsilon)\hat A_k
\right)
\right],
\label{eq:lactor}
\end{equation}
with the probability ratio
\begin{equation}
\eta_k(\bm{\theta})=
\frac{\pi_{\bm{\theta}}(a_k|\mathbf{o}_k)}{\pi_{\bm{\theta}_{\rm old}}(a_k|\mathbf{o}_k)}.
\label{eq:etak}
\end{equation}
The hyperparameter $\epsilon$ sets the clipping range for the policy ratio, which decreases the updating speed and improves the learning stability. The value network is updated by minimizing the mean-squared error,
\begin{equation}
L_{\mathrm{critic}}(\bm{w})
=
\mathbb{E}_k\!\left[
\left(
V_{\bm{w}}(o_k) - \mathcal{G}_k 
\right)^2
\right],
\label{eq:lcritic}
\end{equation}
with the discounted return
\begin{equation}
\mathcal{G}_k = \sum_{l=0}^{N_t-k-1} \gamma^l_{\mathrm{RL}}r_{k+l}.    \label{eq:Gk}
\end{equation}
The total loss minimized during training is
\begin{equation}
L_{\rm total}
=
- L_{\mathrm{actor}}(\bm{\theta})
+
c_1\,L_{\mathrm{critic}}(\bm w)
-
c_2\,S(\pi_{\bm \theta}),
\label{eq:ltot}
\end{equation}
where $c_1,c_2$ are the hyperparameters that balance the three terms, and \(S(\pi_{\bm \theta})\) is the policy entropy that promotes the policy's exploration. Parameters are updated over multiple epochs for each batch of collected trajectories.

In our RL framework, the reward function is designed to  guide the agent to achieve both spin squeezing and stabilization. 
For this purposes, we choose $\xi^2(t)$ and $\xi^2_y(t)$ as the reward metrics, and introduce hyperparameters $\zeta,~\kappa,~\alpha$ to stabilize and accelerate the training process. To discourage the agent from taking unnecessary actions, we include a small penalty term $c(a_k)$ in the reward function. The parameters used for training in the main text are listed in Table~\ref{tab:hyperparameters}.

\begin{table}[t]
\caption{Parameters used for PPO training and environment design.}
\label{tab:hyperparameters}
\centering
\begin{tabular}{p{5.2cm}p{2.2cm}}
\toprule
Parameter & Value \\
\midrule
\multicolumn{2}{c}{\textbf{PPO}} \\
\midrule
Actor learning rate        & \(3\times10^{-4}\) \\
Critic learning rate       & \(1\times10^{-3}\) \\
Discount factor \(\gamma\) & \(0.9999\) \\
GAE parameter \(\hat \lambda\)  & \(0.96\) \\
Clip parameter \(\epsilon\)& \(0.2\) \\
Value-loss weight \(c_1\)  & \(0.5\) \\
Entropy weight \(c_2\)     & \(0.01\) \\
Minibatch size             & 512 \\
Update epochs              & 8 \\
Max train steps            & \(8\times10^6\) \\
Buffer size                & 17920 \\
Actor network              & MLP \((128,128)\) \\
Critic network             & MLP \((128,128)\) \\
\midrule
\multicolumn{2}{c}{\textbf{Environment / reward}} \\
\midrule
spin $f$                         & 21/2\\
Total steps \(N_t\)              & 70 \\
Total time \(\chi T\)            & \(0.314\) \\
Bonus scale \(\zeta\)            & \(5.0\) \\
Decay rate \(\kappa\)            & \(0.05\) \\
Stabilization weight \(\alpha\)  & \(0.05\) \\
Action penalty \(c(a_k)\)        & \(0.001\) \\
\bottomrule
\end{tabular}
\end{table}

\section{Spin stabilization with DD pulses}\label{app:xxx}
As a reference strategy for comparison with the RL protocol, we consider a spin-stabilization scheme which incorporates a dynamic decoupling (DD) sequence optimized by machine learning~\cite{yang2025quantum}. 
In this scheme, the initial state is chosen as the TACT-optimal squeezed state after an additional \(\hat R_x\) rotation such that \(\xi_y^2=\xi^2\). Same as the RL protocol, the encoding time $T_e$ is predefined and discretized with time step $\delta t = T_e/N_e$.
At each discrete time step of duration $\delta t$, the system undergoes an instantaneous rotation about the $x$ axis, followed by QZE evolution. The state update at the \(k\)-th step is
\begin{equation}
|\psi_{k+1}\rangle
= e^{-i\chi dt\,\hat{f}_z^2}\, e^{-i\beta_k \hat{f}_x}\,|\psi_k\rangle ,
\label{eq:updatespx}
\end{equation}
where \(\beta_k\) is the rotation angle applied at the \(k\)th step. 
For a fair comparison with the discrete RL protocol, we use the same time step $\delta t$. The rotation angle $\beta_k$ is arbitrary theoretically, while considering the running speed of the algorithm,  \(\beta_k\) is set to be $n\pi/48$, and the domain is set to $[0,\pi/3]$. In contrast to RL, this scheme is formulated as a conventional parameter optimization problem over the finite set \(\{\beta_k\}\). The optimization is performed using the differential evolution (DE) algorithm~\cite{storn1997differential,lovett2013differential,yang2019improved}. The cost function is defined as the average value of the fixed-axis squeezing parameter during the encoding stage,
\begin{equation}
\mathcal{C}
=
\frac{1}{N_e}\sum_{k=1}^{N_e}\xi_y^2(k),
\label{eq:costspx}
\end{equation}
where \(N_e\) is the total number of discrete steps in the encoding stage.

\section{Phase sensitivity of alternating-$\hat R_y$ pulse sequence}\label{app:phasesensry}

Considering the spin-stabilization pulse sequence consisting of alternating $\hat R_y$ pulses with amplitudes $\pm\beta$ , while the phase $\phi$ are encoded during the free-evolution intervals of duration $\delta t$. For a encoding time $T_e$, the phase accumulates at a rate $\phi/T_e$. One elementary control cycle of duration $dt=2\delta t$ is
\begin{equation}
e^{-i\delta t (\chi \hat f_z^2 +\frac{\phi}{T_e}\hat f_z)} \hat R_y(-\beta) e^{-i\delta t (\chi \hat f_z^2 +\frac{\phi}{T_e}\hat f_z)} \hat R_y(\beta),
\label{eq:cycleencodephase}
\end{equation}
which has the same structure as Eq. (\ref{eq:ucyc}) in the main text, except the phase term $\frac{\phi}{T_e}\hat f_z$. 
Using the relation
\begin{equation}
    \hat R_y(-\beta) \hat f_z \hat R_y(\beta) = \hat f_z \cos \beta - \hat f_x \sin \beta,
    \label{eq:rotated_fz}
\end{equation}
we can rewrite Eq. (\ref{eq:cycleencodephase}) as
\begin{equation}
    e^{-i \delta t \{ \chi[-\hat f_y^2 + (\hat f_{z_\beta}^2 - \hat f_{x_\beta}^2 )\cos \beta] + \frac{\phi}{T_e}\sqrt{2+2\cos \beta}\hat f_{z_\beta} \} +O(\delta t ^2)},
    \label{eq:cycleencodephase_approx}
\end{equation}
with the rotated spin operators
\begin{align}
    \hat f_{z_\beta} &= \hat f_z \cos \frac{\beta}{2} - \hat f_x \sin \frac{\beta}{2}, \\
    \hat f_{x_\beta} &= \hat f_x \cos \frac{\beta}{2} + \hat f_z \sin \frac{\beta}{2}.
    \label{eq:rotspinopers}
\end{align}
As the $\hat f^2_y$ term has a much weaker effect on a spin-squeezed state with small $\Delta f_y$, the phase sensitivity to $\phi$ is main affected by two factors: one is the term $\chi (\hat f_{z_\beta}^2-\hat f_{x_\beta}^2)\cos \beta$ overtwists the state and degrades its squeezing properties; the other one is the phase encoded during the cycle is reduced to $\frac{\phi}{T_e}\sqrt{2+2\cos\beta}\,\delta t$, compared with the real-time value $2\frac{\phi}{T_e}\delta t$. In addition, the encoding axis is changed from $\hat f_z$ to $\hat f_{z_\beta}$. Therefore, before signal encoding, the mean-spin direction should be aligned along the orthogonal axis $\hat f_{x_\beta}$. For a state polarized along the $x$ direction, this corresponds to a rotation $\hat R_y(-\beta/2)$. 

The term $\chi(\hat f_{z_\beta}^2-\hat f_{x_\beta}^2)\cos\beta$ vanishes for $\beta=\pi/2$. In this case, the phase encoded in one cycle becomes $\frac{\phi}{T_e}\sqrt{2}\,\delta t$. Over the encoding time $T_e$, the accumulated phase is $\phi/\sqrt{2}$. Correspondingly, the phase sensitivity is increased as $\Delta\phi=\sqrt{2}\,~\Delta\phi_{\mathrm{std}}$, where $\Delta\phi_{\mathrm{std}}$ is the phase sensitivity in the standard encoding scheme of spin-squeezed states without QZE.
This accounts for the $3.02\,\mathrm{dB}$ reduction of the metrological gain shown in Fig.~\ref{fig:metrology}(a) of the main text.

\begin{figure}
\centering
\includegraphics[width=1.0\columnwidth]{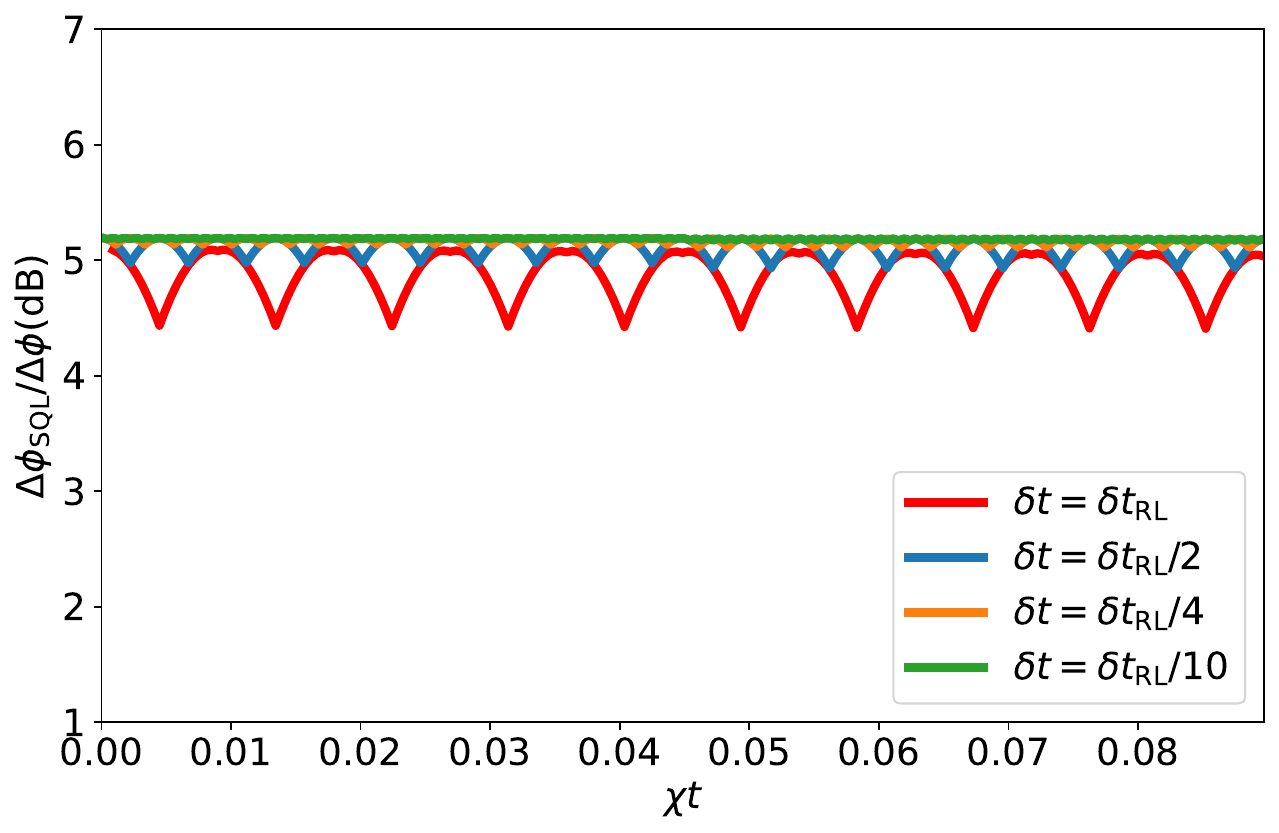}
\caption{\textbf{Phase sensitivity with different pulse intervals $\delta t$.}
The red curve corresponds to the sequence with the same $\delta t$ as that used in the RL protocol in the main text, while the other colored curves correspond to shorter $\delta t$. 
}
\label{fig:appdphivdt}
\end{figure}

Meanwhile, the error term in Eq. (\ref{eq:cycleencodephase_approx}) is the second order in the pulse interval $\delta t$, and the ideal limit is $\delta t\to 0$. Reducing $\delta t$ suppresses the higher-order corrections and makes $\Delta\phi$ more stable, but it cannot overcome the lower bound of $\sqrt{2} \Delta \phi_{\mathrm{std}}$, as shown in Fig.~\ref{fig:appdphivdt}.

\section{Adaptability across different spin-\texorpdfstring{$f$}{f} systems}\label{app:spinf}

\begin{figure}[!ht]
\centering
\includegraphics[width=1\columnwidth]{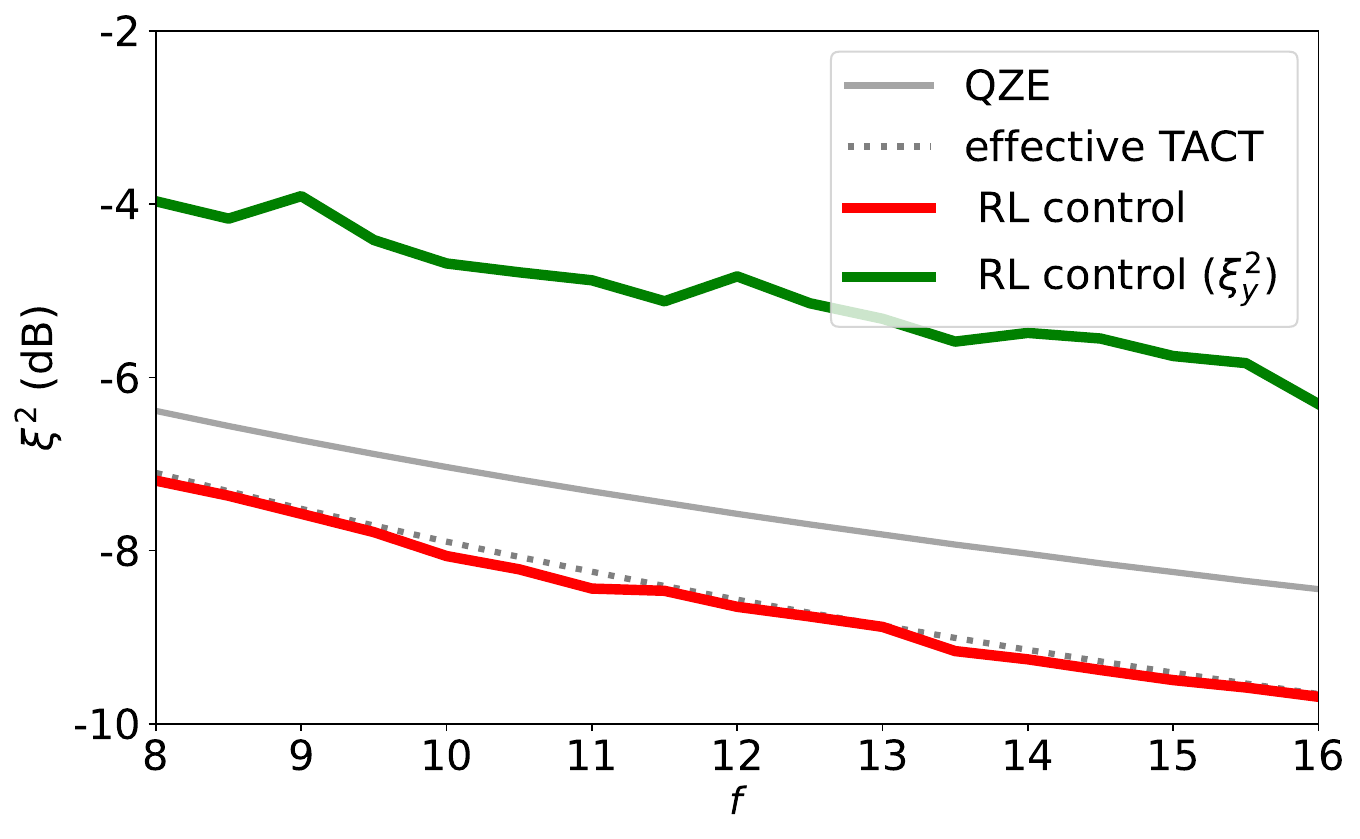}
\caption{\textbf{Train performance across different spin-$f$ systems.}
Squeezing parameters obtained from independently trained RL agents for different spin $f$. 
The red and green curves show the minimum $\xi^2$ reached by the RL protocol and the average stabilized value of $\xi_y^2$. 
The gray solid and dashed curves denote the optimal $\xi^2$ of the QZE and effective TACT models.}
\label{fig:performvf}
\end{figure}

To assess the adaptability of the RL framework, we independently train the agent for a range of spin-$f$ systems and show the squeezing performance in Fig. ~\ref{fig:performvf}. In all cases, the learned protocol exhibits the same qualitative two-stage structure: it first generates squeezing close to the effective TACT benchmark and then stabilizes the fixed-axis squeezing \(\xi_y^2\) within a narrow oscillation window under continued QZE evolution, typically through alternating \(\hat R_y(\pm\pi/2)\) pulses. Quantitatively, the stabilized value of \(\xi_y^2\) decreases overall with increasing \(f\). For a metastable \(f=25/2\) state of Dy~\cite{PhysRevA.50.132}, the protocol stabilizes \(\xi_y^2\) at \(-5.14\,\mathrm{dB}\); for the larger spin \(f=16\), it reaches \(-6.31\,\mathrm{dB}\), approaching the regime relevant to collective-spin squeezing in larger effective-spin systems~\cite{bohnet2016quantum,hines2023spin}. Overall, these results show that the stabilization stage is generally dominated by alternating \(\hat R_y(\pm\pi/2)\) pulses, but the protocols in squeezing stage and the onset of stabilization vary with \(f\), highlighting the advantage of adaptive learning-based control.

\section{Robustness against pulse imperfections}\label{app:pulsenoise}
To assess the experimental robustness of the learned protocol, we consider imperfections in the timing and area of each control pulse. 
For a nominal pulse applied at time $t_k$ with the rotation angle $\beta_k$, we model pulse-timing and pulse-area errors as $t_k \rightarrow t_k+\epsilon_{t,k}\delta t$ and $\beta_k \rightarrow \beta_k+\epsilon_{\beta,k}\beta_k$, respectively, where $\delta t$ is the control interval in the main text.
The dimensionless fluctuations $\epsilon_{t,k}$ and $\epsilon_{\beta,k}$ are independently sampled from Gaussian distributions, $\epsilon_{t,k} \sim \mathcal{N}(0,\sigma_t^2)$ and $\epsilon_{\beta,k} \sim\mathcal{N}(0,\sigma_{\beta}^2)$.
These imperfections are included throughout both the state-preparation and sensing stages.

For each pair of $(\sigma_t,\sigma_\beta)$, we simulate the complete protocol over 500 independent realizations and evaluate the averaged magnetic-field sensitivity 
\begin{equation}
    \overline{\delta B} = \frac{1}{N_{\mathrm{r}}}
\sum_{j=1}^{N_{\mathrm{r}}}
\delta B_j,
\end{equation}
where $N_{\mathrm{r}}$ denotes the number of realizations and $\delta B_j$ is the magnetic-field sensitivity obtained from the $j$th realization.

\begin{figure*}
\centering
\includegraphics[width=0.98\linewidth]{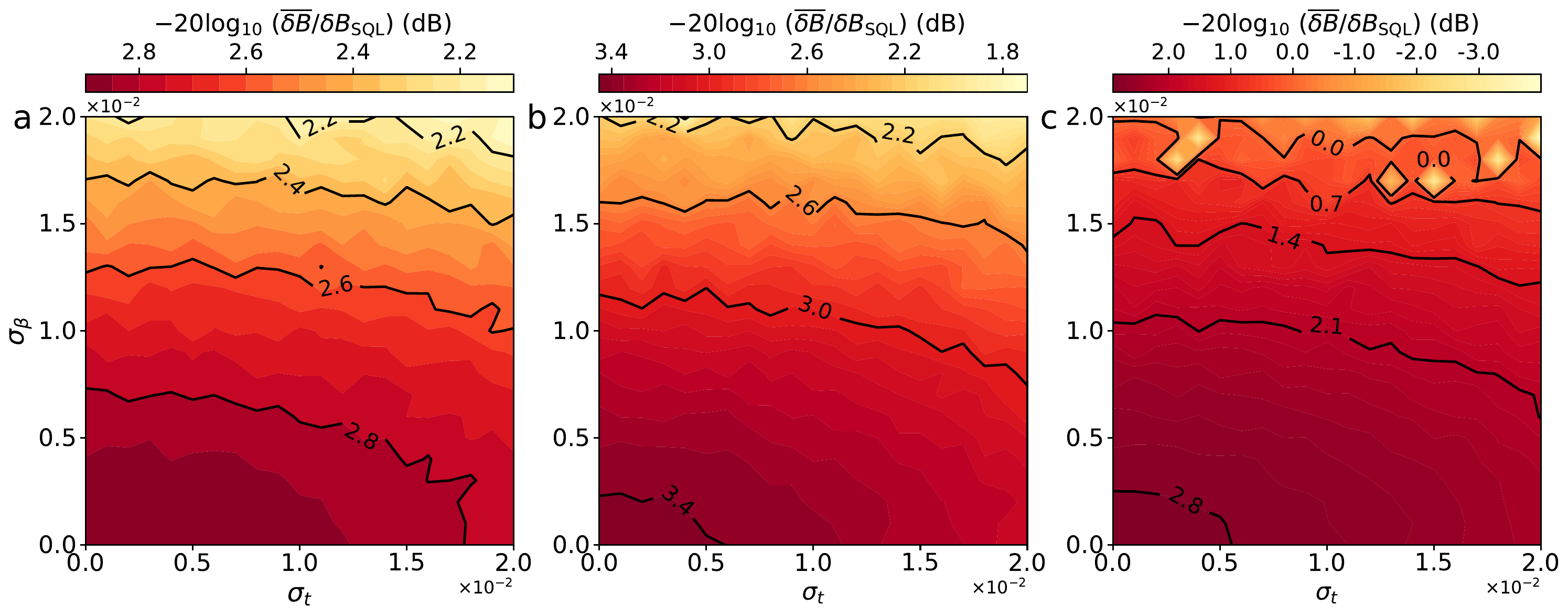}
\caption{\textbf{Robustness of the RL protocol against pulse imperfections.}
The SQL-relative magnetic-field sensitivity $\delta B_{\mathrm{SQL}}/\delta B$ as a function of pulse-timing and pulse-area errors $(\sigma_t,\sigma_\beta)$, averaged over 500 realizations with independent Gaussian perturbations for different total protocol times $T_{\mathrm{tot}}$.(a)   $T_{\mathrm{tot}} = 10\, \mathrm{ms}$ . (b) $T_{\mathrm{tot}} = 17.5\, \mathrm{ms}$. (c) $T_{\mathrm{tot}} = 30\, \mathrm{ms}$. }
\label{fig:appdpulsenoise}
\end{figure*}

Fig.~\ref{fig:appdpulsenoise} shows the SQL-relative magnetic-field sensitivity $\delta B_{\mathrm{SQL}}/\overline{\delta B}$ as a function of $\sigma_t$ and $\sigma_\beta$ for three representative total protocol times.
The RL protocol is less sensitive to pulse-timing errors than to pulse-area errors over the parameter range considered.
In addition, the degradation becomes more pronounced as \(T_{\rm tot}\) increases, consistent with the accumulation of pulse imperfections over the larger number of control operations required for longer sensing times. 
Nevertheless, a beyond-SQL sensitivity persists over a broad region of the parameter space, including simultaneous timing and pulse-area errors at the percent level.

\section{Effect of dephasing on magnetic-field sensitivity}\label{app:dephasing}
The analysis in the main text assumes unitary dynamics.
To assess the influence of decoherence on the long-time metrological performance, we introduce a dephasing channel associated with fluctuations along the quantization axis.
We re-simulate the system dynamics while retaining the RL pulse sequence optimized under unitary dynamics.
The evolution is governed by the Lindblad master equation
\begin{equation}
    \frac{d\rho}{dt} = -i[\hat{H}(t),\rho]
    + \Gamma_z \left( \hat f_z\rho\hat f_z - \frac{1}{2} \left\{\hat f_z^2,\rho
\right\} \right),
\label{eq:masterequation}
\end{equation}
where $\Gamma_z$ characterizes the dephasing strength.
The dephasing channel is included throughout both the state-preparation and encoding stages. 

\begin{figure}[!ht]
\centering
\includegraphics[width=1.0\linewidth]{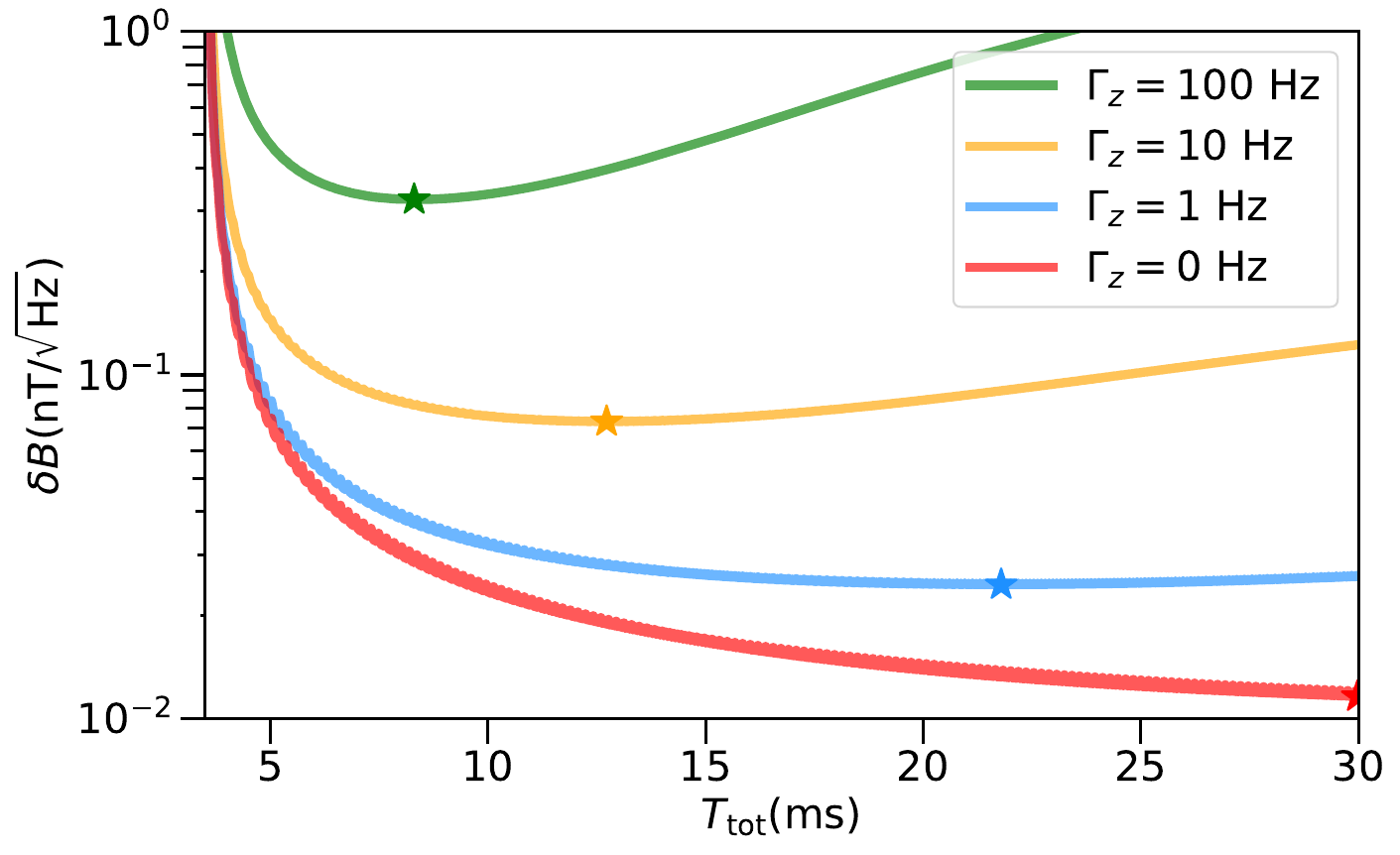}
\caption{\textbf{Effect of dephasing on the RL protocol.}
The magnetic-field sensitivity $\delta B$ as a function of the total protocol time $T_{\mathrm{tot}}$ for dephasing strengths $\Gamma_z=0, 1, 10$ and $100\,\mathrm{Hz}$.
Stars mark the minimum sensitivity for each $\Gamma_z$.
}
\label{fig：dephasing}
\end{figure}

Fig~\ref{fig：dephasing} shows the magnetic-field sensitivity $\delta B$ as a function of the total protocol time $T_{\mathrm{tot}}$ for several dephasing strengths $\Gamma_z=0,1,10,$ and $100\,\mathrm{Hz}$.
In the absence of dephasing, the sensitivity continues to improve over the time range considered, as in Fig.~\ref{fig:metrology}(c).
However, the longer-time improvement is no longer sustained once dephasing is included.
Instead, the sensitivity reaches an optimum at a finite total protocol time and subsequently degrades.
As $\Gamma_z$ increases, the optimal protocol time shifts to shorter values.
These results show that the achievable sensitivity is ultimately limited by decoherence, while sub-$\mathrm{nT}/\sqrt{\mathrm{Hz}}$ sensitivity can be achieved for dephasing strengths on the order of tens of hertz.
With improved experimental control and further suppression of magnetic field noise, longer coherence times would allow the protocol to maintain its metrological advantage for longer times.

\bibliography{mybib}

\end{document}